# Charge-Neutral Electronic Excitations in Quantum Insulators

Sanfeng Wu[1*], Leslie M. Schoop[2], Inti Sodemann[3], Roderich Moessner[4], Robert J. Cava[2], N. P. Ong[1*]

[1] Department of Physics, Princeton University, Princeton, New Jersey 08544, USA
[2] Department of Chemistry, Princeton University, Princeton, New Jersey 08544, USA
[3] Institute for Theoretical Physics, University of Leipzig, 04103 Leipzig, Germany
[4] Max-Planck Institute for the Physics of Complex Systems, 01187 Dresden, Germany

*Email: sanfengw@princeton.edu; npo@princeton.edu

## Abstract

Experiments on quantum materials have uncovered many interesting quantum phases ranging from superconductivity to a variety of topological quantum matter including the recently observed fractional quantum anomalous Hall insulators. The findings have come in parallel with the development of approaches to probe the rich excitations inherent in such systems. In contrast to observing electrically charged excitations, the detection of charge-neutral electronic excitations in condensed matter remains difficult, though they are essential to understanding a large class of strongly correlated phases. Low-energy neutral excitations are especially important in characterizing unconventional phases featuring electron fractionalization, such as quantum spin liquids, spin ices, and insulators with neutral Fermi surfaces. In this perspective, we discuss searches for neutral fermionic, bosonic, or anyonic excitations in unconventional insulators, highlighting theoretical and experimental progress in probing excitonic insulators, new quantum spin liquid candidates and emergent correlated insulators based on two-dimensional layered crystals and moiré materials. We outline the promises and challenges in probing and utilizing quantum insulators, and discuss exciting new opportunities for future advancements offered by ideas rooted in next-generation quantum materials, devices, and experimental schemes.

## Main

In crystalline solids, electrons are packed closely together and interact strongly via the Coulomb force. In the standard Landau paradigm, the electronic quasiparticle excitations are viewed as dressed weakly interacting fermions although they couple to external electric and magnetic fields via their electric charge. This powerful description - valid in familiar metals, semiconductors, and insulators - has worked remarkably well in condensed matter physics. It underlies virtually all modern electronics-based technologies from transistors, solar cells, and light emitting diodes to superconducting devices.

However, materials that do not obey this conventional description do exist. An example is the situation of interacting electrons confined to a one-dimensional (1D) wire, where the Luttinger liquid (LL) theory, rather than Landau's Fermi liquid theory, serves as the standard description.[1] The low-energy excitations in a LL are fractionalized modes which separately carry spin and charge degrees of freedom. In 2D and 3D systems, non-Fermi liquid physics has increasingly occupied center stage in research on strongly correlated materials, including cuprates[2], heavy fermion materials[3,4], and more recently moiré materials[5–8]. Beyond non-Fermi liquids, strongly correlated electrical insulators may also exhibit striking properties that go beyond the conventional picture. This is the theme of our perspective. We review how the conventional quantum theory of solids may break down in electrical insulators, discuss the current experimental status and outline promising future directions. We focus on insulators exhibiting interesting charge-neutral electronic excitations in the bulk that can't be captured by conventional band theory.

## Theoretical Overview of Neutral Excitations & Fractionalization in Insulators

Up until the 1970's it was widely believed that excitations in many-electron systems would always carry an integer multiple of the electron charge, and that particles with an even (odd) multiple of electron charge would necessarily be bosons (fermions). Charge neutral excitations like excitons (bound electron-hole pairs) or magnons (spin excitations in a magnet) were expected to be bosons. The discoveries of fractionalization in 1D polymer chains[9,10] and the fractional quantum Hall (FQH) effect in 2D electron gases[11] disrupted this paradigm by demonstrating that materials can harbor quasiparticles with a fractional electron charge (**Fig. 1a**). Moreover, quasiparticles in fractionalized electronic systems can have anyonic exchange statistics[12–14] and are therefore fundamentally differed from bosons or fermions.

Separately, Anderson proposed his seminal ideas of the resonant valence bond (RVB) "spin liquid" states[15], which gained much prominence as proposed parent states of high-$T_c$ superconductors[16]. The RVB proposal led to strong debate. A model exhibiting a bona fide RVB spin liquid phase was found a few decades later[17]. The RVB spin liquid and the FQH states are now understood to be examples of a large class of fractionalized states[18]. For example, the short-range RVB liquid in the absence of symmetry represents the same topological phase[18] as the toric code introduced by Kitaev[19]. These are often referred to as $Z_2$ spin liquids on account of the local Ising ($Z_2$) symmetry of the emergent gauge field.

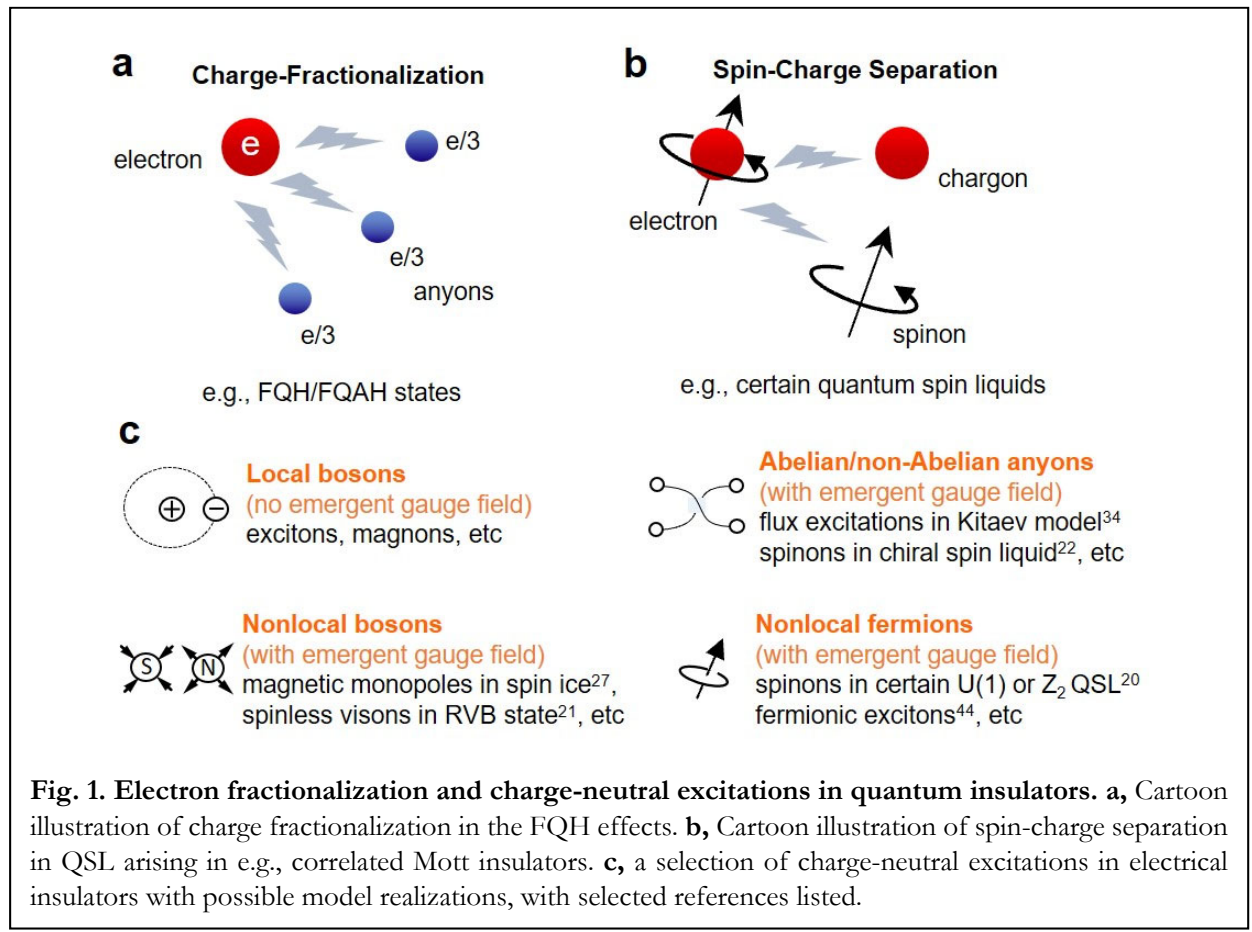

**Fig. 1. Electron fractionalization and charge-neutral excitations in quantum insulators. a,** Cartoon illustration of charge fractionalization in the FQH effects. **b,** Cartoon illustration of spin-charge separation in QSL arising in e.g., correlated Mott insulators. **c,** a selection of charge-neutral excitations in electrical insulators with possible model realizations, with selected references listed.

Indeed, the presence of *emergent* deconfined gauge fields is often considered to be the defining feature of a spin liquid. The emergent gauge field in spin liquids comes with accompanying particles that are charged under it. The deconfinement is the property that these particles can be separated arbitrarily far away from each other with a finite energy cost. This endows them with a non-local aspect: they are sources/sinks of a corresponding emergent flux, just as an electron is a source of electric field on account of its electric charge. These particles, neutral with respect to Maxwell electromagnetism, can have different quantum statistics: fermionic, bosonic or anyonic. There certainly is no need for all neutral particles to be bosons. Quite spectacularly, they can carry quantum numbers that are *fractions* of those of the microscopic degrees of freedom[20–22]: for example, in an RVB liquid, adding an electron can generate two *independent* particles: a spinon that carries the spin and a chargon that carries the electric charge (**Fig. 1b**).

Mathematically, the case of an emergent U(1) gauge field is very close to Maxwell electromagnetism, which is also a U(1) gauge theory. However, details can differ greatly: the emergent theory can harbor both electric and magnetic emergent charges, and exhibit a much larger fine-structure constant[23]. One can even study the emergent analogue of Maxwell theory in two dimensions. However, this is believed to be stable only when the gauge field coexists with gapless fermionic degrees of freedom charged under this gauge field.[24,25] These fermions could be nodal (as in e.g. Dirac spin liquids) or have a Fermi surface (as in e.g. the "spinon Fermi surface state"). In three dimensions the U(1) gauge structure can remain stable even in the absence of gapless fermionic degrees of freedom. A model system known as spin ice, a paradigmatic model displaying an emergent U(1) gauge structure in 3D,[26,27] features emergent electric and magnetic charges[28]. **Figure 1c** summarizes properties and occurrences of selected charge-neutral quasiparticles in insulators.

The electromagnetic response of correlated insulators with fractionalized excitations is rich and complex. Local probe fields can only create multiplets of fractionalized quasiparticles, as the net quantum numbers of the product of a scattering process cannot be fractional. Unlike the case of neutrons scattering off single magnons, it is therefore generally not possible to obtain a sharp response even from long-lived fractionalized quasiparticles. The coupling matrix elements can further differ hugely from those of conventional electromagnetism, not least on account of different symmetry properties of emergent fields; for instance, emergent magnetic fields can be even under time reversal. Nonetheless, U(1) spin liquids with gapless fermions can in principle display striking experimental consequences, including subgap power law optical conductivity[29], quantum oscillations in response to magnetic fields[30,31], cyclotron resonance in the absence of charged quasiparticles[32], and metallic-like magnetic noise[33], etc. All of these provide new avenues for their experimental detection. See more discussions and their implications on measurable quantities in **Box 1**.

**Box 1 Fractionalization and Emergent Gauge Fields in Insulators**

***An Exactly Soluble Model*** – The Kitaev's honeycomb model[34] is exactly soluble and may also be relevant for experiments on candidate materials. Its Hamiltonian, in the simplest incarnation, reads

$$H_K = \sum_{ij,\alpha} \sigma_i^\alpha \sigma_j^\alpha ,$$

where the sum is over all nearest neighbor pairs *ij*. It is 'spin-orbit coupled' in that the spins only interact via one component of their spin (given by Pauli matrix $\sigma^\alpha$, $\alpha = x, y, z$) depending on the bond direction, as depicted in **Fig. 3d**.

An elegant way to demonstrate fractionalization is to directly identify the degrees of freedom carrying the fractional quantum numbers. We write the spin-1/2 operators as a combination of four Majorana operators, three 'bond Majoranas' $b_i^\alpha$ and one 'matter Majorana' $c_i$. Majorana operators have commutation relations $\{b_i, b_j\} = 2\delta_{ij}$ and are often referred to as 'real fermions', as a pair of them can be combined to a standard complex fermion, $f$, as $f^\dagger = (c_1 - ic_2)/2$. The model has a set of non-dynamical (i.e., conserved and immobile) flux operators, $W_p = \prod_o \sigma_i^\gamma$, one for each hexagon plaquette "*p*" of the honeycomb lattice. Here, the product defining $W_p$ runs over the six sites of *p*, with $\gamma$ the bond direction pointing out of the hexagon at site *i*. Upon writing $S_i^\alpha = b_i^\alpha c_i$, one obtains that the $W_p$ operators are entirely made from the 'bond Majoranas' $b_i^\alpha$. The values of all $W_p$ can independently take $\pm 1$ (i.e., they are $Z_2$ degrees of freedom), and together they define a flux sector. The Kitaev honeycomb model can then be exactly solved, flux sector by flux sector, by considering a fermionic hopping problem for the matter fermions ($f$ and $f^\dagger$) subject to the background fluxes $W_p$.

*Fractional Excitations* - This solubility illustrates explicitly several general features of fractionalization: the microscopic spin degrees of freedom have been replaced by more 'natural' emergent variables, the matter fermions and fluxes. These emergent variables capture the 'breaking apart' of the spin degrees of freedom. In principle, one obtains such a form of the Hamiltonian from more general parton constructions, whose utility depends on how naturally the chosen construction reflects the actual low-energy degrees of freedom of the many-body problem under consideration. When the itinerant fermions ($f$ and $f^\dagger$) have a gapped spectrum with Chern number $\pm 1$, the Kitaev model realizes a non-abelian fractionalized state. There are two kinds of quasiparticles in the bulk of this non-abelian state: the complex itinerant fermions ($f$ and $f^\dagger$) and the Ising flux-like particle (the vison) associated with the above plaquette operator. A vison is present (absent) in a plaquette when $W_p = -1$ ($W_p = +1$). The vison particle, which is an analogue of an Abrikosov vortex in a superconductor, carries a Majorana zero mode in its core and exhibits non-abelian exchange statistics. **Fig. 1c** summarizes neutral excitations in selected models.

*Responses* - How does a physical electromagnetic field couple to the emergent degrees of freedom? The answer in general is clear: it depends. On the level of the microscopic Hamiltonian, a simple Zeeman field $h^\alpha$ in the Kitaev model couples to both matter fermions and fluxes: $h^\alpha S_i^\alpha = i h^\alpha b_i^\alpha c_i$. Its action on the fractionalized system can hence be quite involved – e.g., it can give dynamics to the flux degrees of freedom, while also changing the matter degrees of freedom. This illustrates why the coupling to electromagnetic fields is so diverse in fractionalized systems: simple-looking couplings turn out to be quite complex once the spins have broken apart.

Indeed, each fractionalized model in principle comes with its own coupling to external fields, which will depend on symmetry properties of the emergent degrees of freedom as well as underlying microscopic details. For example, in the case of U(1) QSL featuring a spinon Fermi surface, Landau quantization may develop as a response of the spinon-gauge field system to an external magnetic field[30]. In quantum spin ice, fractionalization leads to an emergent gauge field which has photons, electric charges (spinons) but also – unlike the world we inhabit – magnetic charges (**Fig. 1c**). How an externally applied electromagnetic field couples to these depends on microscopic features, such as: do the spins derive from ions with an even or an odd number of electrons? In the latter case, time-reversal symmetry, as expressed e.g. in Kramers theorem, forbids certain couplings of an electric field to the fractionalized excitations.

Generally, the Hamiltonian of the fractionalized particles involves not only the emergent gauge coupling, but also 'remembers' properties such as the crystal field scheme of the microscopic spins. Determining many features in detail is a challenge for each material individually. Nonetheless, simple behavior can emerge from qualitative considerations. In the case of spin ice, the fractionalized quasiparticles are sources of both an emergent field as well as the usual (Maxwell) magnetic field—hence their appellation as magnetic monopoles. Such a combination has been termed a hybrid dyon[35], and it can lead to effects, superficially familiar from metallic systems, appearing in novel ways in insulators. An interesting prediction is a magnetic Nernst effect in insulators, in which applying an electric field perpendicular to a temperature gradient induces a magnetization perpendicular to both.[35]

## Experimental Progress, Challenges & Opportunities

Experimentally, the detection of quantum properties of insulators and charge-neutral excitations is hampered by the lack of suitable low-temperature probes. Thermal transport is one of the few techniques available for investigating excitations in insulators at low $T$ in strong magnetic fields. While it is a powerful probe for 3D materials, it faces challenges in 2D materials, a growing platform for many interesting insulating phases. Other conventional approaches, such as electrical transport and tunneling spectroscopy, are not directly sensitive to charge-neutral excitations, especially when the charge gap is large. This limitation may explain why experimental progress in 2D insulators has been slow except in insulators with a gap < ~ 100 meV. Next we discuss the status and perspective for the experimental detection of hidden quantum phenomena in insulators, focusing on selected topics in both 3D bulk materials and the rapidly evolving correlated 2D crystals and van der Waals (vdW) stacks.

### Case I – Excitonic Insulators

The first prominent case we highlight is the excitonic insulator (EI). In the single-particle band picture, we have a band insulator if a fully occupied valence band is separated from the conduction band by a finite energy gap $E_g$. This picture is modified when the gap is small because Coulomb interaction leads to bound states (excitons) of electrons in the conduction band and holes in the valence band[36–38], as sketched in **Fig. 2a**. This effect occurs when the exciton binding energy $E_{ex}$ exceeds the band gap $E_g$. A similar situation is obtained in low carrier-density semimetals in the limit of weak screening of the Coulomb interaction. The formation of excitons converts the semimetal to an insulator, as pointed out by Mott[39]. Theories of excitonic anomalies near the semimetal-semiconductor transition are reviewed in detail by Halperin and Rice.[40]

In brief, in weakly interacting band insulators there is an energy gap to create excitons. However, as the electron repulsions increase this gap can decrease leading to quantum phase transitions into a new state where excitons spontaneously "proliferate", without a closing of the charge gap. When the exciton is an ordinary local boson (see **Fig.1c**) the resulting state would be an insulator with a spontaneously broken symmetry[40]. In this case the EI can represent a special subset of broken symmetry states where the exciton condensation is additionally accompanied by some robust form of nearly gapless fluidity of neutral modes, which remains robust throughout the phase of matter itself and not just near a critical point. One mechanism for this is that the exciton carries a quantum number associated with a continuous symmetry, then its condensation leads to the existence of broken symmetry (quasi)Goldstone modes. Another mechanism is that the exciton carries a momentum which is incommensurate with the reciprocal Bravais vectors, and the broken symmetry state resulting from the exciton condensation will be an incommensurate charge density wave, spin density wave [40] or spin spiral state[41,42], whose sliding modes will remain partly soft and cannot be completely pinned by the atomic crystal. There can also be more exotic forms of exciton "proliferation"[43,44], which are not necessarily local bosons and can lead to protected gapless fluid neutral modes, such as the composite exciton fermion[44] (**Fig. 1c**).

Experimental confirmation of an excitonic insulator in crystals is challenging because the excitons are intrinsic to the ground state and are charge-neutral. Note that exciton physics in optically excited semiconductors is well established and has been extensively studied in the past decades in 2D semiconducting transition metal dichalcogenides (TMD)[45,46]. One can straightforwardly observe signatures of optical excitons in photoluminescence measurements since these excitons can decay by emitting light. In EIs, excitons are the lowest energy state and their observation is not straightforward.

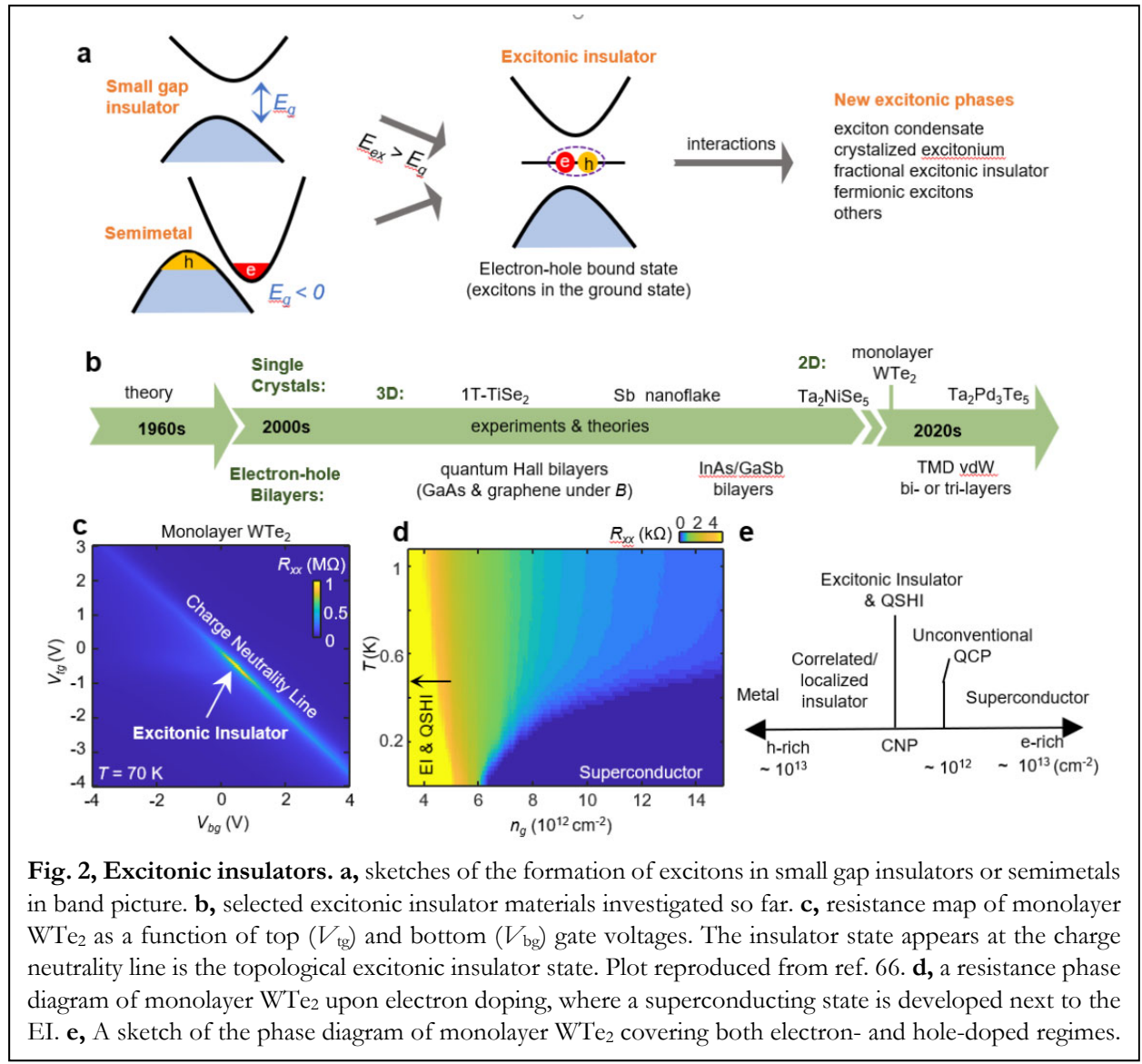

**Fig. 2, Excitonic insulators. a,** sketches of the formation of excitons in small gap insulators or semimetals in band picture. **b,** selected excitonic insulator materials investigated so far. **c,** resistance map of monolayer WTe$_2$ as a function of top ($V_{tg}$) and bottom ($V_{bg}$) gate voltages. The insulator state appears at the charge neutrality line is the topological excitonic insulator state. Plot reproduced from ref. 66. **d,** a resistance phase diagram of monolayer WTe$_2$ upon electron doping, where a superconducting state is developed next to the EI. **e,** A sketch of the phase diagram of monolayer WTe$_2$ covering both electron- and hole-doped regimes.

Experimental investigation of EIs fall into two categories: (1) *artificial electron-hole bilayers* – Investigation of exciton condensates has been systematically conducted in quantum Hall bilayers consisting of two closely spaced 2D electron gasses placed in high magnetic fields[47–49]. Striking consequences of the quantum Hall exciton condensate include Josephson-like quantum tunneling, Coulomb drag and counterflow transport, etc, as observed in both semiconductor quantum wells and graphene[47–49]. Evidence for an EI state in zero magnetic field has also been reported in InAs/GaSb bilayers[50,51] and more recently in several van der Waals bilayers or trilayers of TMDs[52–54] (**Fig. 2b**). (2) *Natural crystals* – While the EI theory was originally developed for bulk crystals, its experimental detection has been more challenging. Candidate bulk crystals include 1T-TiSe$_2$[55–57], Sb nanoflakes[58], Ta$_2$NiSe$_5$[59–63] and recently Ta$_2$Pd$_3$Te$_5$[64,65] (**Fig. 2b**). A common challenge in identifying an EI phase in bulk crystals is to distinguish it from a trivial band insulator especially since structural transformations in these systems generate ongoing debates (e.g., ref[63]). In 2021, a 2D crystal, i.e., monolayer WTe$_2$, was identified as an EI[66,67]. **Fig. 2c** plots the resistance of monolayer WTe$_2$ as a function of the top and bottom gate voltages typically employed for 2D devices, highlighting the appearance of the EI state at charge neutrality. The gate tunability of a 2D crystal allows for new opportunities in characterizing an EI. In particular, the gate-tuned tunneling spectra[66] of monolayer WTe$_2$ reveal that the insulator phase originates from electron correlations and rule out the possibility of a band insulator. This key conclusion is further supported by gate-tuned measurements of the chemical potential[67] and an anomaly in the gate-tuned Hall effect[66]. Such gate-tuned measurements are infeasible in bulk crystals.

2D WTe$_2$ therefore serves as an exciting test bed for investigating characteristics of EIs[41,42,66–68]. The promise is further highlighted by the many other properties of this material, including (1) the EI is also a quantum spin Hall insulator (QSHI)[69–72], (2) a superconducting phase emerge when a moderate density of electrons is introduced to the EI[73,74] (**Fig. 2d**), (3) a distinct intriguing phase[66,67,70,72,75] on the hole doped side (**Fig. 2e**), and (4) unexpected Landau quantization that appears in the insulating regime[75,76] (see discussions in case III below). Moreover, the recent finding of an unconventional quantum critical point[77] located in between the EI phase and the superconducting phase highlights a deeper connection between the two to be discovered. The topological EI state at charge neutrality, as the pristine property of the monolayer, is core to unlock the many intriguing mysteries associated with the rich low-$T$ phenomena in this material.

The field of EI has gained new impetus and more EI materials are emerging. In general, interactions in the many-exciton state of an EI are likely to generate a diverse variety of phases. The major challenge is that our experimental approaches for investigating a charge neutral state are immature such that examinations of even the plain vanilla version of an EI remain difficult.

### *Case II - Quantum Spin Liquids*

The quantum spin liquid (QSL) state was proposed by Anderson in 1973[15] as a new kind of insulator that exhibits a liquid-like ground state of spins on a lattice with geometric frustration (**Fig. 3a**). Subsequent theoretical work has shown that the QSL may host exotic fractional excitations and the emergent gauge fields. One remarkable possibility is that there is a class of gapless QSL featuring a spinon Fermi surface accompanied with an emergent U(1) gauge field. While it is true that the spinons do not carry charge under the physical magnetic field, the internal gauge field may couple to the external magnetic fields. For instance, the emergent U(1) gauge field structure may allow for the appearance of an emergent magnetic field whose average value itself can be induced by an external magnetic field in some situations[30,31,78], producing spinon Landau quantization. Several recent reviews on QSLs summarize both the theoretical and experimental advances to date[79–83]. Despite much progress, however, the existence of a QSL state in a physical material remains an open question.

We first highlight the experimental challenges on this topic by tracing the studies of one specific QSL candidate 1T-TaS$_2$, a vdW layered material named in Anderson's original paper[15]. This material was not widely recognized as a QSL candidate until 2017, when Law and Lee provided arguments[84] empowered by modern views. A possible spinon Fermi surface state in 1T-TaS$_2$ was later proposed.[85] At low $T$, the material develops an insulating charge density wave consisting of clusters of stars of David, forming a triangular superlattice (**Fig. 3b**). Experimental studies on the nature of this state, in principle a Mott insulator, are challenging. Investigations of its bulk form involves the interlayer stacking order, which complicates the situation[86–90]. Studies on monolayer 1T-TaS$_2$, a much stronger insulator compared to its bulk, can clear out the situation but little is known about the monolayer. Recently, scanning tunneling microscope studies of monolayer 1T-TaSe$_2$ (a sister material of 1T-TaS$_2$) grown on metallic graphene substrates (**Fig. 3c**), necessary for charge transport, have shown interesting results consistent with the Mottness and possible QSL state.[91–93] However, the low-$T$ properties of the pristine monolayers of 1T-TaS$_2$ or 1T-TaSe$_2$, either isolated or on an insulating substrate, remains unknown due to the lack of proper experimental tools to diagnose such a strong 2D insulator. Similar studies and status can be found in the isostructural 1T-NbSe$_2$.[94–97]

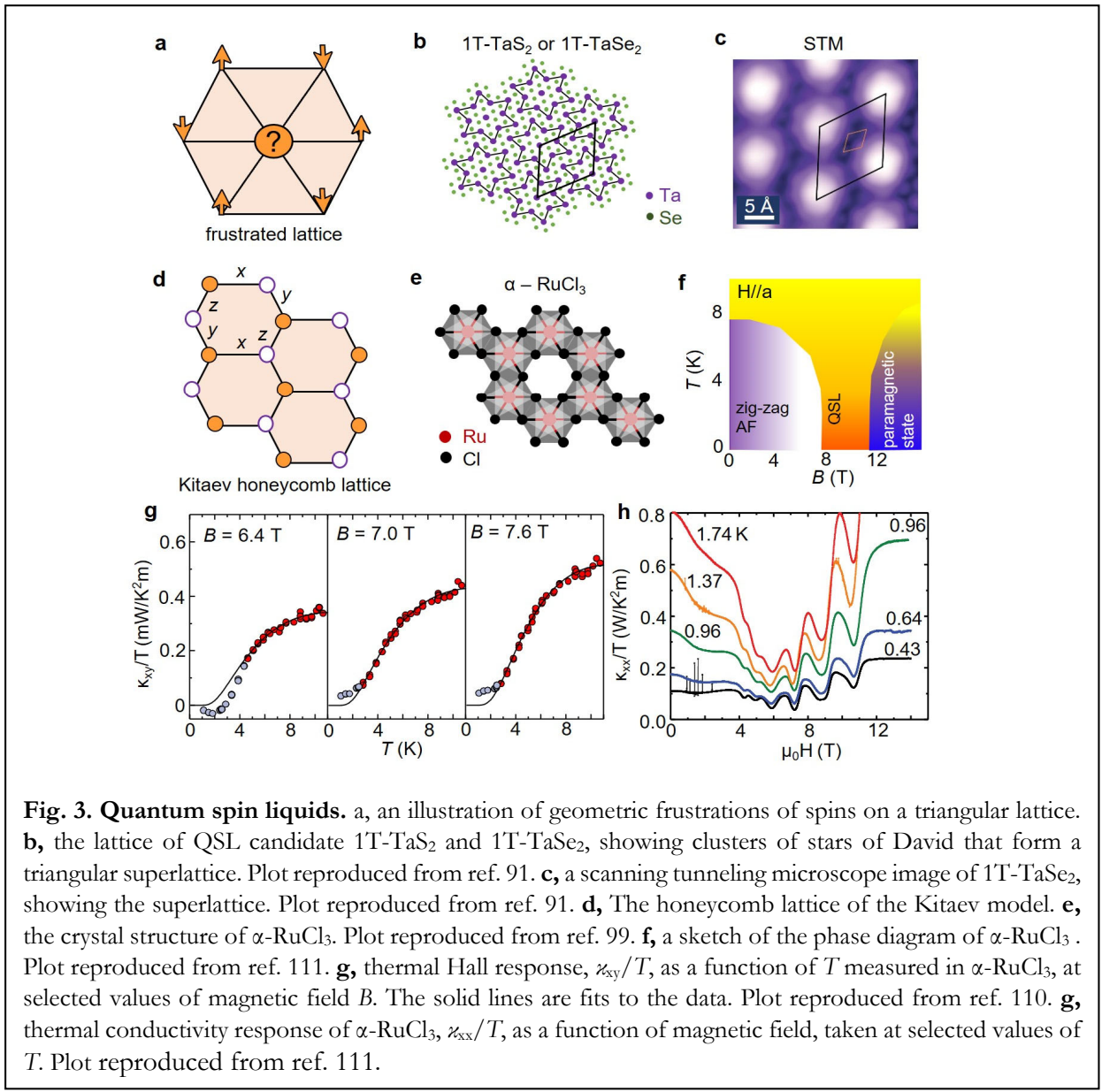

**Fig. 3. Quantum spin liquids.** **a,** an illustration of geometric frustrations of spins on a triangular lattice. **b,** the lattice of QSL candidate 1T-TaS$_2$ and 1T-TaSe$_2$, showing clusters of stars of David that form a triangular superlattice. Plot reproduced from ref. 91. **c,** a scanning tunneling microscope image of 1T-TaSe$_2$, showing the superlattice. Plot reproduced from ref. 91. **d,** The honeycomb lattice of the Kitaev model. **e,** the crystal structure of α-RuCl$_3$. Plot reproduced from ref. 99. **f,** a sketch of the phase diagram of α-RuCl$_3$. Plot reproduced from ref. 111. **g,** thermal Hall response, $\kappa_{xy}/T$, as a function of $T$ measured in α-RuCl$_3$, at selected values of magnetic field $B$. The solid lines are fits to the data. Plot reproduced from ref. 110. **g,** thermal conductivity response of α-RuCl$_3$, $\kappa_{xx}/T$, as a function of magnetic field, taken at selected values of $T$. Plot reproduced from ref. 111.

Another topical area of QSL research is the search for materials that are proximate to the Kitaev honeycomb model Hamiltonian[34] $H_K$ which describes spin-1/2 Ising spins interacting with bond-specific exchange couplings (see **Box 1** and **Fig.3d**)[34,82,98]. Following calculations[98] pointing to the 4$d$ and 5$f$ chalcogenides as likely platforms, α-RuCl$_3$ (**Fig. 3e**) was identified as the closest proximate[99]. The Ru ions, which carry a spin-1/2 moment, occupy the sites of the honeycomb lattice. In zero magnetic field, α-RuCl$_3$ orders at 7 K as an antiferromagnet (with zig-zag order stabilized by terms not included in $H_K$). An in-plane field **H,** applied parallel to the zig-zag axis **a**, causes long-range order to vanish sharply at a critical field $B_{c2}$ = 7.3 T. As **H** is increased above ~10 T, sharp magnon modes emerge in the spectrum. The interesting state in-between (7.3 T < $H$ < 10 T), characterized by a broad, featureless spectrum of excitations, has been widely discussed as a QSL state relatable to $H_K$[99–107] (**Fig. 3f**). Both the spin liquid and zig-zag states have been investigated by neutron scattering[99], electron spin resonance[108], terahertz spectroscopy[109], thermal transport[110,111], etc.

Interest spiked following a report that, within a narrow temperature window (3.7 and 4.9 K), the thermal Hall conductivity $\kappa_{xy}$ in the QSL interval seemed to exhibit[112] a half-quantized value in accord with Kitaev's calculation, both with **H** tilted out of plane[112] and with **H∥a** (the zigzag axis)[113]. In this scenario, the heat current is carried by Majorana excitations that occupy chiral edge modes. However, subsequent experiments have not validated the finding of half-quantization. Measurements extended to a much broader interval in $T$ (0.5 to 10 K, with **H∥a**) reveal[110] that $\kappa_{xy}$ arises from excitations that obey the Bose-Einstein distribution. In the proposed picture, the spin excitations occupy a topological band, and are subject to a large Berry curvature that changes sign with **H∥a**.[114] Fits to the Murakami expression for $\kappa_{xy}$ yield a Chern number (of the lowest spin band) close to 1 above 9 T (**Fig. 3g**). Moreover, the inferred band energy is ~1 meV, in good agreement with previous microwave absorption and electron spin resonance experiments. Subsequent experiments comparing crystals grown under different conditions again do not observe half-quantization. There is one report[115] that argues that $\kappa_{xy}$ may approach the half-quantized value for $H$ larger than 10 T. In this regime, however, the uncertainties in $\kappa_{xy}$ diverge uncontrollably because the thermal Hall resistivity $\lambda_{yx}$ rapidly vanishes[110].

Another finding in α-RuCl$_3$ is that, below 4 K, the thermal conductivity $\kappa_{xx}$ exhibits large oscillations vs. $H$ (with **H∥a**)[111] (**Fig. 3h**). Although α-RuCl$_3$ is an excellent charge insulator, the $\kappa_{xx}$ curves resemble Shubnikov-de Haas (SdH) oscillations in a semimetal. The integers $n$ indexing the extrema in $\kappa_{xx}$ vary linearly with $1/H$, except for a sharp break in slope near $B_{c2}$ = 7.3 T. The oscillation amplitudes are strongly enhanced within the QSL field interval (7.3, 10) T although a tail of attenuated oscillations persists to 4 T, deep in the zig-zag state. In the 4 samples studied[111], the oscillations are closely similar both in phase and period, but the amplitudes are sample dependent. An interesting interpretation is that, despite the complete absence of free electrons, long-lived neutral fermionic excitations appear to define an effective Fermi surface. These fermions have been proposed to be pseudo-scalar spinons[116] in order to reconcile the presence of oscillations of $\kappa_{xx}$ concomitant with absence of $\kappa_{xy}$ when the in-plane field is along **b** (the armchair axis). Note that Landau quantization is not expected for the original Kitaev model. Alternately, the oscillations are modulations of the dominant phonon conductance caused by periodic oscillations of the spin-density of states which modulate the scattering amplitude between phonons and spin excitations.

Subsequently, oscillations in $\kappa_{xx}$ vs. $H$ were observed by Bruin et al.[117] in crystals of $\alpha$-RuCl$_3$ displaying a large spread of $T_N$'s from 7 to 13-14 K, which they attributed to a cascade of stacking faults. Originally, Kubota et al.[118] had proposed that sweeping $H$ at $T$ <4 K induces multiple transitions caused by stacking-fault (SF) creation. Kubota et al linked multiple features in the field profile of the magnetic susceptibility at 4 K to the large spread of $T_N$ by angular extrapolation within the $T$-$H$ plane. Later, Cao et al.[119] showed that the spread in $T_N$'s (especially the 14 K transition in the magnetization) in a given crystal provides a reliable signature of high SF density (this test is now widely adopted to screen crystals with high SF density). Bruin et al.[117] adopted Kubota's cascading scenario to identify the oscillations with successive transitions by field-creation of SFs at low $T$ (see also Lefrancois et al.[120]). Zhang et al.[121,122] have tested this association by monitoring the oscillations in a series of crystals in which the SF density is controlled. They found that the oscillation period is actually closely similar in all crystals irrespective of SF density, with amplitude largest in crystals with lowest SF density. They conclude that the oscillations are unrelated to the 14-K transition. We note that none of the crystals used in Czajka et al[111] exhibits the 14-K transition or even a spread of $T_N$'s. Together, these studies strongly disfavor the cascading transition scenario. We further remark that the stochastic nature of field-induced SF should lead to widely different oscillation periods (and hysteresis) in conflict with all reports to date. A recent experiment[123] on the dependence of the thermal conductivity *vs.* the azimuthal angle of the in-plane **H** reveals an angular variation consistent with an intrinsic property of the QSL state. Future experiments examining other quantities beyond thermal conductivity may provide new insights into the nature of this intriguing state. More recently, Hong et al.[124] reported $\kappa_{xx}$ v.s. $H$ measured in the Kitaev material Na$_2$Co$_2$TeO$_6$, which displays two prominent dips at low $T$ for both field orientations (**H** along armchair or zigzag axes). Hong et al.[124] propose that these features, which roughly resemble the oscillations reported by Czajka et al in $\alpha$-RuCl$_3$, arise from phonon scattering from magnetic structures or domains in the spin-disordered state in Na$_2$Co$_2$TeO$_6$.

### Case III - The Search for Charge-Neutral Fermi Surfaces in Insulators

The quest to find insulators that exhibit a neutral Fermi surface (e.g., QSL with a spinon Fermi surface) has long been a challenge in condensed matter physics. Unlike in conventional insulators, the thermal conductivity here is predicted to exhibit a metallic temperature profile at very low temperatures. This is one of the key tests[79–81] so far adopted in many experiments, e.g., on organic materials[79] and 3D quantum magnets[125]. Currently, there is considerable debate on the results and their interpretations (see e.g., ref[126–128]). A central feature of the spinon Fermi surface is the emergent gauge field which couples to spinons in a way similar to the coupling of the electromagnetic fields to electrons. This leads to observable effects if the emergent gauge field further couples to electromagnetic fields.

A useful concept governing the electromagnetic response of a fractionalized system is the Ioffe–Larkin rule[129,130]. For example, electronic transport in a spin-charge separated material involves both spinon transport and chargon transport. However, the source and drain metal electrodes can only inject/accept electrons, which fractionalize into spinons and chargons within the material. By the Ioffe-Larkin rule, the measured resistivity (not conductivity) is the sum of the spinon resistivity and chargon resistivity. An intriguing case is the Landau quantization of the spinon Fermi surface in magnetic fields[30], which can in principle lead to quantum oscillations (QOs) of observables in insulators [30,31,131]. Experimentally, the QOs look similar to the SdH and de Haas-van Alphen effects in metals.

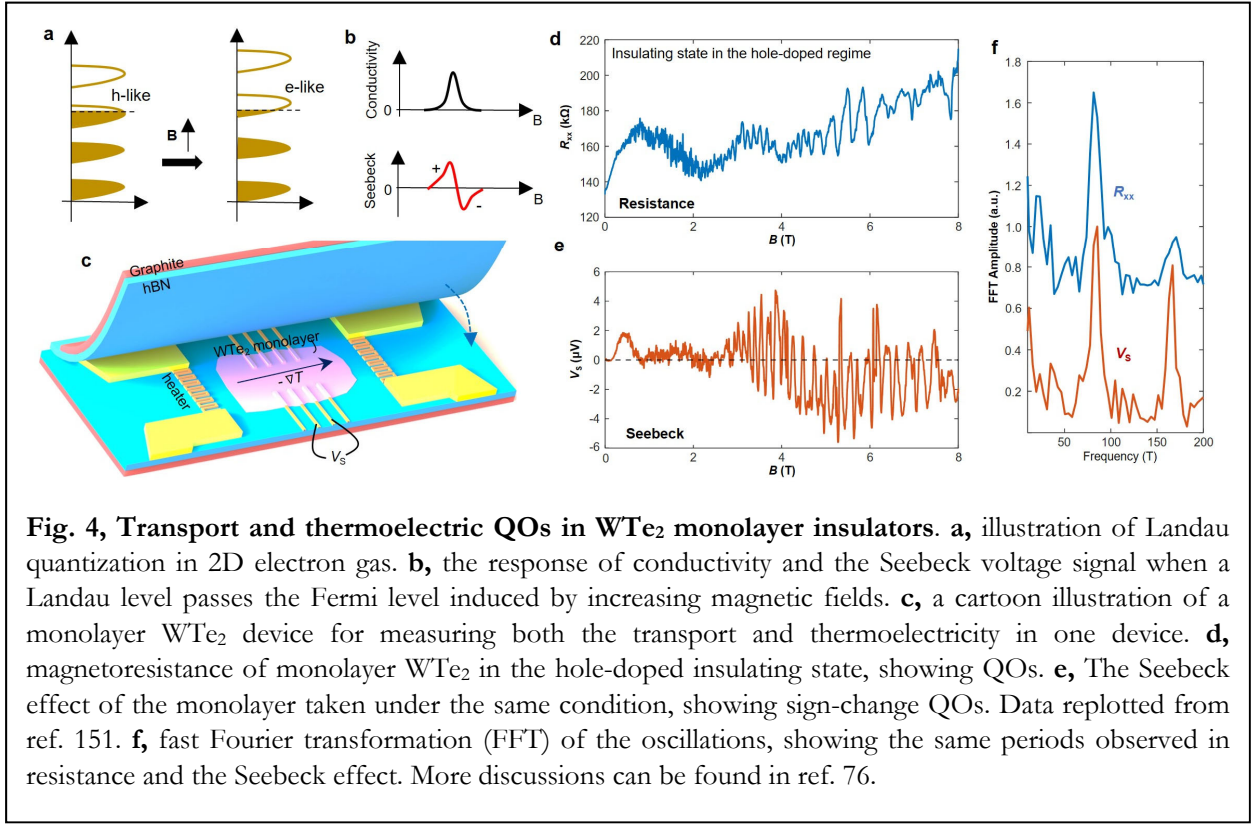

**Fig. 4, Transport and thermoelectric QOs in WTe$_2$ monolayer insulators**. **a,** illustration of Landau quantization in 2D electron gas. **b,** the response of conductivity and the Seebeck voltage signal when a Landau level passes the Fermi level induced by increasing magnetic fields. **c,** a cartoon illustration of a monolayer WTe$_2$ device for measuring both the transport and thermoelectricity in one device. **d,** magnetoresistance of monolayer WTe$_2$ in the hole-doped insulating state, showing QOs. **e,** The Seebeck effect of the monolayer taken under the same condition, showing sign-change QOs. Data replotted from ref. 151. **f,** fast Fourier transformation (FFT) of the oscillations, showing the same periods observed in resistance and the Seebeck effect. More discussions can be found in ref. 76.

Although QOs in insulators have been reported in several systems including Kondo insulators SmB$_6$[132,133] and YbB$_{12}$[134], Quantum Wells[135,136], topological EI WTe$_2$[75], the QSL candidate α-RuCl$_3$[111], and more recently on YCu$_3$-Br[137], the discussion of their interpretations remains widely open[117,120,123,138–150]. In the previous section, we summarized the status of oscillations observed in α-RuCl$_3$. A conclusive demonstration of a neutral Fermi surface inside a charge gap remains an outstanding goal. It is essential to establish a concrete case where intrinsic Landau quantization in insulators can be firmly established experimentally. This requires a combination of techniques for detecting QOs that simultaneously excludes competing explanations.

We highlight the promise of achieving such a goal in monolayer WTe$_2$ and subsequently strongly correlated 2D materials systems. QOs in monolayer WTe$_2$ insulators were first reported in the resistance measurements[75]. The most striking observation lies in the apparent conflict between the measured low conductivity (σ) of the resistive state (> 100 MΩ) and the high carrier density ($n > 10^{12}$ cm$^{-2}$) with high mobility (μ >1,000 cm$^2$ V$^{-1}$s$^{-1}$) extracted from the QO data[75]. In other words, σ << $n e$ μ. A natural interpretation to resolve this conflict is to assign the highly mobile carriers responsible for the QOs to be charge neutral. This is the neutral fermion picture. In alternative explanations, these highly mobile particles are ordinary charge carriers, either thermally activated or residing in the metal component used in the device. In one scenario (*i*), they are charge carriers thermally activated across the gap, and the QOs of the insulator are interpreted as the consequence of the *B*-induced oscillations of the insulator gap[68,145]. In a second scenario (*ii*), the highly mobile carriers reside in the nearby graphite gate, while carriers in monolayer WTe$_2$ merely serve to detect Landau levels in graphite[146].

To distinguish these scenarios, one needs to go beyond electrical transport measurements. In particular, measurements of QOs in the thermoelectric response of monolayer WTe$_2$ have considerably clarified the situation[76,151]. Distinct from the conductivity, the Seebeck effect is sensitive to the derivative of the density of states with respect to energy and hence to the energy structure near the chemical potential. For instance, in a Landau quantized 2D electron gas, the Seebeck signal will develop a sign-change oscillation each time when the Fermi level is crossed by a Landau level (see e.g., observations in graphene[152,153]), signifying that the carrier type has altered its character effectively from "hole-like" to "electron-like" (**Figs. 4a & b**). Such sign-change thermoelectric QOs have been observed recently in the WTe$_2$ monolayer insulator in the hole doped regime[76] (**Figs. 4c-f**), implying that the highly mobile carriers responsible for the QOs belong to WTe$_2$ insulator itself and that a Landau-level like energy structure is developed in magnetic fields near the chemical potential. The alternative scenarios (i) and (ii) are not supported by the thermoelectric data. In the neutral Fermi surface scenario, the Ioffe-Larkin sum rule relates the physical thermopower to the behaviors of all fractionalized components. The thermopower of neutral fermions, defined in relation to the emergent electric field, may be sensitive to the Landau quantization of the fermions[129,130]. An exact formulation for such a spin-charge separated insulator with Landau quantization remains to be developed. The results identify 2D WTe$_2$ as an intriguing platform for investigating unconventional insulators.

More experimental probes are desirable to extend investigation of the QOs in 2D and 3D quantum insulators. The efforts towards confirming the true nature of insulating WTe$_2$ and its puzzling Landan quantization problem will lead to the development of a new set of tools for diagnosing 2D insulators. It is exciting that we now have several concrete 2D material platforms, including WTe$_2$, 1T-TaS$_2$/1T-TaSe$_2$, and α-RuCl$_3$, to explore the possibilities of neutral fermions, charge neutral fermi surface and other neutral phases in insulators. More candidate materials will surely emerge in the future.

## The Quest for New Materials, New Experiments and New Theories

What is necessary to advance the research includes the developments of new quantum insulators, new detection schemes, innovative device structures and a new theoretical understanding of next-generation quantum effects in insulators. Such endeavors require synergetic efforts between chemists, experimentalists, and theorists.

***New Materials*** – New synthesis of quantum materials in both the bulk and 2D forms is crucial. Recent examples of new Kitaev materials include Ru-based[154] (e.g., RuI$_3$) and Co-based[155–157] honeycomb lattice materials (e.g., BaCo$_2$(AsO$_4$)$_2$). 2D forms of quantum insulators are still rarely explored and hold great promise, yet we already have several target materials including but not limited to the several cases that we highlighted in this article. The unique advantages of gate turnabilities and interface effects, especially via moiré quantum engineering, in 2D vdW crystals may provide entirely new possibilities in addressing some key issues. For instance, beyond monolayers, twisted bilayers of QSL may allow for quantum engineering of spinon bands.

***New Detection Schemes*** – The detection of neutral excitations in condensed matter requires innovative techniques beyond conventional means. For instance, while we have no issues with fabricating high-quality devices of monolayer and twisted bilayer 1T-TaS$_2$, we don't have an approach to uncover its electronic properties at low *T*. Recently, far-infrared optical spectroscopy of quantum materials at millikelvin temperatures and in magnetic fields has been developed[158], which might be helpful moving forward. Optical means can in principle access neutral excitations that are not visible in charge transport. For example, optical resonances of ground state excitons may provide an unambiguous demonstration of excitonic insulators and distinguish their species. Fingerprints of a charge neutral Fermi surface may be found in sub-gap optical resonances[29,32,78,159]. Similarly, probing techniques in the THz/GHz regime will likely be powerful, yet their applications in correlated quantum insulators, especially at ultralow *T* and in strong *B*, remain largely unexplored. A promising avenue is to extract information of hidden excitations by two-dimensional coherent spectroscopy[160–162]. Another exciting direction to explore is to employ recent advances in quantum sensing for probing quantum matter. A proposal is to utilize NV-center spin qubits for detecting quantum noises of spinon Fermi surfaces[33,163–165].

***Charge-Neutral Quantum Devices*** – The investigations of neutral quantum phases in insulator will likely benefit the development of future quantum devices. The exploration of such devices will in turn further promote the study of quantum insulators. For instance, neutral excitonic or spinon phases may allow for low-power consumption transistors without suffering Joule heating (as it doesn't involve charge transport). Simultaneously, chip-scale generation, nonlocal transport[166] and detection of neutral excitations could provide unique ways of proving the existence of hidden neutral modes. Such explorations will be essential in establishing the unique nonlocal properties of many neutral excitations of interest. A highly relevant topic is that the realization of electron fractionalization and anyons in insulators can in principle enable robust quantum computation schemes with topological protection[14].

***New Theoretical Understanding*** – Despite the significant progress in theoretical understanding of a large class of non-trivial quantum insulators we still face serious challenges in bridging the divide between ideal theories, models, and materials. One strategy for closing this divide is to search for new general principles that could enrich our search for the emergence of fractionalization in models and materials. For example, non-trivial band topology prevents the existence of a trivial localized flat-band Hubbard limit, because of the obstruction to constructing Wannier orbitals. This offers an enhancement of the quantum fluctuations induced by interactions that could favor non-trivial quantum orders in partially filled topological bands. In fact, this feature is intimately linked to why fractionalized phases are abundant in the limit of Landau levels, which are essentially ideal flat Chern bands. While the link of non-trivial band topology to enhanced quantum fluctuations has been recently emphasized and investigated in moiré materials, understanding its broader implications and incarnations, for example in 3D materials, remains largely open. With the increasing numbers of 2D and 3D insulators with non-trivial topological band structures, the need to better understand the interplay of band-topology and strong interactions is pressing.

Another important challenge is to develop the theory of novel experimental probes that help identify and characterize the presence of non-trivial states in materials. For example, NV center noise spectroscopy is a promising tool that could help overcome some of the difficulties of NMR in 2D settings, and could help identify different non-trivial states.[33,163–165] More generally there is a need to develop our understanding of probes that could guide better the detection and characterization of non-trivial states in correlated materials.

## Summary and outlook

In this perspective, we have provided our views on a class of intriguing problems of quantum insulators and reviewed their status in theory and experiment. The topics discussed are selected to emphasize what we believe are the major challenges and promises in the field. Space restrictions prevent inclusion of many other recent findings. An example is the fractional quantum anomalous Hall (FQAH) effect[167–172] in fractional Chern insulators[173–177] recently observed in moiré materials. The rapid pace of discovery suggests that research in charge-neutral phases, excitations, and phase transitions in fractionalized moiré materials will be an exciting area in the next few years.

The challenges in detecting hidden phenomena in insulators seem to recall how electromagnetic waves were first detected. We have made analogies between neutral fermions coupled to the emergent gauge field and electrons coupled to electromagnetic fields. The Hertz experiments[178], performed two decades after publication of Maxwell's equations, transformed the communication industry. Success in detecting neutral excitations and emergent gauge fields in quantum insulators may have similar impact.

## Acknowledgements
S. W. acknowledges support from Gordon and Betty Moore Foundation's EPiQS Initiative Grant GBMF11946, AFOSR through a Young Investigator Award (FA9550-23-1-0140), ONR through a Young Investigator Award (N00014-21-1-2804), NSF through a CAREER award (DMR-1942942), and the Sloan Foundation. N. P. O., R. J. C., L. M. S. and S.W. acknowledge support from the Materials Research Science and Engineering Center (MRSEC) program of the NSF (DMR-2011750). N.P.O. acknowledges support from the United States Department of Energy (DE-SC0017863) (which supported the $\kappa_{xy}$ experiments on $\alpha$-RuCl$_3$) and the Gordon and Betty Moore Foundation through Grant GBMF9466. L.M.S. acknowledges support from the Gordon and Betty Moore Foundation through Grants GBMF9064 and the David and Lucile Packard Foundation. S.W. and L.M.S. acknowledge support from the Eric and Wendy Schmidt Transformative Technology Fund at Princeton. The research on forefront electronic materials in the laboratory of R. J. C. at Princeton is supported by the Gordon and Betty Moore Foundation through grant GBMF-9066 and the US DOE division of Basic Energy Sciences (DE-FG02-98R45706). I.S. acknowledges support from the Deutsche Forschungsgemeinschaft (DFG) through research grant project numbers 542614019 and 518372354. R.M. acknowledges funding by the Deutsche Forschungsgemeinschaft under grants SFB 1143 (project-id 247310070) and the cluster of excellence ct.qmat (EXC 2147, project-id 390858490).

## Author contributions
S.W. and N.P.O initiated the work. I.S. and R.M. wrote the theory parts. S.W., N.P.O., R.J.C., and L.M.S. wrote the experimental parts. All authors discussed and contributed to the overall writing and revisions of this work.

## Competing interests
The authors declare no competing interests